\pdfoutput=1
\RequirePackage{scrlfile}
\PreventPackageFromLoading{natbib}
\PassOptionsToPackage{final}{graphics}
\PassOptionsToPackage{unicode,pdfencoding=auto}{hyperref}

\documentclass[english,3p]{elsarticle}
\usepackage{ifdraft}

\usepackage{ifxetex}
\usepackage{ifluatex}
\usepackage{ifthen}
\usepackage[final]{microtype}[2011/08/18]

\usepackage{fixltx2e}
\usepackage{textcase}
\usepackage{fontaxes}
\usepackage{relsize}
\usepackage{etoolbox}
\ifthenelse{\boolean{xetex}}{
  \usepackage{fontspec}
  \newfontfeature{Microtype}{protrusion=default;expansion=default;}
  \defaultfontfeatures{}
  \setmainfont[ExternalLocation,
    Microtype,Ligatures=TeX,
    Extension=.otf,
    UprightFont= *-regular,
    BoldFont=*-bold,
    ItalicFont=*-italic,
    BoldItalicFont=*-bolditalic]{texgyretermes}
  \setsansfont[ExternalLocation,
    Microtype,Ligatures=TeX,
    Scale=MatchLowercase,
    Extension=.otf,
    UprightFont= *-regular,
    BoldFont=*-bold,
    ItalicFont=*-italic,
    BoldItalicFont=*-bolditalic]{texgyreheros}
  \usepackage{unicode-math}
\unimathsetup{math-style=TeX}
  \setmathfont{xits-math.otf}

  \microtypesetup{protrusion}
  \let\textls\undefined
  \DeclareRobustCommand{\textls}[2][80]{%
    {{\addfontfeature{LetterSpace=\the\numexpr#1/10\relax}#2}}}
}{
  \usepackage[LGR,T1]{fontenc}
  \usepackage{txfonts}
  \usepackage[utf8]{inputenc}
  \usepackage{alphabeta}
  \microtypesetup{protrusion,stretch=9,shrink=15,step=3,letterspace=25,tracking=smallcaps}
  \DeclareFontFamily{T1}{pzc}{}
  \DeclareFontShape{T1}{pzc}{m}{it}{<-> s * [0.900] pzcmi7t}{}
  \DeclareMathAlphabet{\mathscr}{T1}{pzc}{m}{it}
}

\usepackage{color}
\usepackage[table,svgnames]{xcolor}

\usepackage{fixltx2e}
\usepackage{graphicx}
\usepackage[export]{adjustbox}
\usepackage{ifthen}
\usepackage{tabularx}
\usepackage{fancyvrb}
\usepackage{xspace}
\usepackage{relsize}
\usepackage{setspace}
\usepackage{booktabs}
\usepackage{pifont}
\usepackage[binary-units]{siunitx}
\usepackage{acronym}
\usepackage{rotating}
\usepackage[shortcuts]{extdash}
\ifxetex

\else

\fi

\usepackage[final]{listings}

\newcommand\hairspace{\ifmmode\mskip1mu\else\kern0.08em\fi}
\newcommand{\eg}{for example\xspace}

\newcommand{\ie}{that is\xspace}

\newcommand{\cf}[1]{cf.\,{#1}\xspace}

\makeatletter
\newcounter{algorithm}
\renewcommand\thealgorithm{\@arabic\c@algorithm}
\newcommand*\algorithmname{Algorithm}
\def\fps@algorithm{tbp}
\def\ftype@algorithm{4}
\def\ext@algorithm{lox}
\def\fnum@algorithm{\algorithmname\nobreakspace\thealgorithm}
\newenvironment{algorithm}
               {\@float{algorithm}}
               {\end@float}
\newenvironment{algorithm*}
               {\@dblfloat{algorithm}}
               {\end@dblfloat}
\makeatother

\usepackage{wrapfig}
\newcommand{\id}[1]{\ensuremath{\textit{#1}}}

\newcommand{\prop}[2]{{#1}{\def\tempa{#2}%
\ifmmode\mathsmaller{\{\tempa\}}%
\else\textsmaller{\tempa}\fi}}
 
\usepackage[%
  final, pdfpagelabels=false,
  hidelinks, %
  plainpages=false,%
]{hyperref}

\newif\ifblind
\blindfalse
\makeatletter
\let\els@c@author\c@author
\let\c@author\relax

\let\bibfont\relax
\makeatother

\usepackage[%
natbib=true,
backend=biber,%
style=elsevier-num,
doi=true,url=true,%
isbn=false,%
]{biblatex}

\iffalse %
\bibliography{bibliography}
\else
\fi

\newcommand*\LambTitle{Adaptive Just-in-time Value Class Optimization}
\newcommand*\LambSubtitle{for Lowering Memory Consumption and Improving
  Execution Time Performance}
\newcommand*\LambKeywords{Meta-tracing \sep JIT
  \sep Data Structure Optimization \sep Value Classes}

\edef\LambPdftitle{\LambTitle{}%
\ifx\LambSubtitle\empty\relax%
\else \LambSubtitle%
\fi}
\hypersetup{
  pdfkeywords={\LambKeywords},
  pdftitle={\LambPdftitle},%
}

\ifblind\else
\hypersetup{
    pdfauthor={%
      Tobias Pape, %
      Carl Friedrich Bolz, %
      Robert Hirschfeld%
}}
\fi

\journal{Science of Computer Programming}

\begin{document}
\hyphenation{RPy-thon tech-nique Be-be-ni-ta Has-kell dem-on-strat-ed
  prom-is-ing Fa-kul-tät}
\newacro{JIT}[\textsc{jit}]{just-in-time}
\newacro{AOT}[\textsc{aot}]{ahead-of-time}
\newacro{AST}[\textsc{ast}]{abstract syntax tree}
\newacro{CIL}[\textsc{cil}]{Common Intermediate Language}
\newcommand\RPython{R\texorpdfstring{\kern.03em}{}Py\-thon\xspace}
\newcommand*\shape[2][]{%
  \def\tempa{#1}%
  \ifx\tempa\empty%
  \textsf{\textit{#2}}%
  \else%
  \textsf{\textit{#2\textsubscript{\tempa}}}%
  \fi%
}
\newcommand*\constr[1]{%
  \textsf{\textsmaller{#1}}\xspace%
}

\makeatletter
\let\bib@c@author\c@author
\let\c@author\els@c@author
\makeatother


\begin{frontmatter}

\title{\LambTitle{}\penalty-100 \LambSubtitle{}}

\ifblind
\author{(removed for blind review)}
\else

\author[hpi]{Tobias Pape\corref{c}}
\ead{tobias.pape@hpi.uni-potsdam.de}
\author[kings]{Carl Friedrich Bolz}
\ead{cfbolz@gmx.de}
\author[hpi]{Robert Hirschfeld}
\ead{hirschfeld@hpi.uni-potsdam.de}
\cortext[c]{Corresponding author}

\address[hpi]{Software Architecture Group, Hasso Plattner Institute, University of Potsdam, Germany}
\address[kings]{Software Development Team, King's College London, UK}

\fi


\begin{abstract}
  The performance of value classes is highly dependent on how they are
  represented in the virtual machine. Value class instances are immutable, have
  no identity, and can only refer to other value objects or primitive values
  and since they should be very lightweight and fast, it is important to
  optimize them carefully. In this paper we present a technique to detect and
  compress common patterns of value class usage to improve memory usage and
  performance. The technique identifies patterns of frequent value object
  references and introduces abbreviated forms for them. This allows to store
  multiple inter-referenced value objects in an inlined memory representation,
  reducing the overhead stemming from meta-data and object references. Applied
  to a small prototype and an implementation of the Racket language, we found
  improvements in memory usage and execution time for several micro-benchmarks.
\end{abstract}


\end{frontmatter}

\makeatletter
\let\els@c@author\c@author
\let\c@author\bib@c@author
\makeatother

\section{Introduction}

The way data structures are represented affects their performance. Especially
virtual machine developers carefully choose the representation of their data
structures, classes, or objects so that using them is efficient. In this paper
we propose, implement, and evaluate an optimized representation for \emph{value
  classes}~\cite{bacon:2003:kava:-java} on the virtual machine level. Value
class instances are immutable objects without identity that can reference only
other value classes instances or primitive data. They have been suggested for a
an extended Java~\cite{bacon:2003:kava:-java}, Java
itself~\cite{rose:2014:169:-value}, exist in .NET~\cite{msdn:2014:common-type}
and \--- in a limited form \--- in Scala~\cite{odersky+:2006:overview-scala}. However,
related constructs of immutable identity-less structures also occur in several
other languages, particularly in functional ones. Examples include the
algebraic data types of ML and Haskell, Prolog's terms, cons cells in certain
LISPs\footnote{This is a special case, since LISP only supports one ``value
  type'', \texttt{cons}. Also, other LISPs exist where cons cells do have
  identity or are mutable.}, and structures in Racket~\cite{plt-tr1}.
Therefore, our optimization should be applicable to a number of other contexts.
Nevertheless, in this paper we will use the terminology \emph{value classes}
and \emph{instances of value classes} (\emph{value objects} for short).

The simplest approach to a machine representation of value objects is a class
pointer together with their fields as a list of pointers to other value objects
and primitive values. We propose an object layout that stores nested value
object groups in a compacted, linearized fashion. This works by observing that
in practice some shapes in the object graph are much more common than other
shapes. There are often repeating patterns of how value objects reference each
other. For example, a cons cell is likely to reference another cons cell in its
tail field, or a tree node often references other tree nodes.

For such common shapes we \emph{inline} the fields of the referenced value
object into the referring object to save space and to accelerate the traversal
of the object graph. This inlining can be repeated with fields of nested value
objects, potentially several levels deep. We detect which object graph shapes
are common by keeping statistics at run-time, since it is often impossible to
statically infer what shapes will be common in practice.\footnote{Note that
  these shapes are totally different what some JavaScript VMs such as Firefox'
  IonMonkey and Higgs call shapes. Those JavaScript ``shapes'' are equivalent
  to Self \emph{maps} or V8's hidden classes. We will discuss the relationship
  to Self maps in the related work section.} The inlining is only possible
because of the key properties of value objects:
\begin{enumerate}[a)]
\item Value objects are immutable, so the reference to an inlined object can
  never be replaced by another reference.
\item Value objects do not have identity, so the fact that an inlined object
  does not have a separate memory address that can be used as its identity does
  not create problems. Likewise, multiple copies of an inlined object are not
  problematic for identity concerns.
\end{enumerate}

We implement the proposed optimization in two prototypes. One implements a
variant of the lambda calculus extended with value objects and pattern matching,
which we used to prototype and evaluate the proposed optimization in isolation.
To also evaluate the approach in a more realistic setting, we implemented the
same optimization for Pycket~\cite{spenser-bauman+:2015:pycket:-tracing}, a
re-implementation of the Racket language. Both languages use the RPython virtual
machine implementation framework and its tracing \acf{JIT} compiler. The
tracing \ac{JIT} compiler is instrumental to our approach since it is
responsible for producing fast machine code for accessing the modified
representation.

The contributions of this paper are as follows:
\begin{itemize}
\item We propose an approach for finding patterns in value object usage at run-time.
\item We present a compressed layout for value objects that makes use of those
    patterns to store value objects more efficiently.
\item We report on the performance of micro-benchmarks for a small prototype
  language and a Racket implementation.
\end{itemize}

The paper is structured as follows: \autoref{sec:tracing-jit} gives a brief
introduction to tracing \acs{JIT} compilers. In \autoref{sec:optim-appr}, we
present our approach to just-in-time optimization of data structures. Our two
implementations are presented briefly in \autoref{sec:impl} and their
performance is evaluated in \autoref{sec:results}. Our approach is put into
context in \autoref{sec:related-work} and we conclude in
\autoref{sec:future-directions}.



\section{Tracing Just-In-\kern-0.12emTime Compilers}
\label{sec:tracing-jit}

We briefly introduce tracing \acf{JIT} compilers~\cite{bolz:2012:meta-tracing-just-in-time}, as some of their properties
are key to the performance characteristics of our approach
(cf.~\autoref{sec:compr-through-inlin} and s~\autoref{sec:impl-field-access}).

Just-in-time (\acs{JIT}\acused{JIT}) compilation has become a mainstream
technique for, among other reasons, speeding up the execution of programs at
run-time. After its first application to LISP in the 1960s, many other language implementations
have benefitted from \ac{JIT} compilers---from APL, Fortran, or
Smalltalk and Self~\cite{aycock:2003:brief-history} to today's popular
languages such as Java~\cite{paleczny+:2001:java-hotspot} or
JavaScript~\cite{holtta:2013:crankshafting-from}.

One approach to writing \ac{JIT} compilers is using \emph{tracing}~\cite{bala+:2000:dynamo:-transparent}. A tracing
\ac{JIT} compiler records the steps an interpreter takes in common execution
paths such as hot loops. The obtained instruction sequence is commonly called a
\emph{trace}. This trace can on be optimized independently or transformed to machine
code and used instead of the interpreter to execute the same part of that
program~\cite{mitchell:1970:design-construction} at higher speed. Tracing
produces specialized instruction sequences, \eg for one path in if--then--else
constructs; if execution takes a different branch later, it switches back to use the interpreter. Tracing \ac{JIT}
compilers have been successfully used for optimizing native code%
~\cite{bala+:2000:dynamo:-transparent} and also for efficiently executing
object-oriented programs~\cite{gal+:2006:hotpathvm:-effective}.

\emph{Meta-tracing} takes this approach one step further by observing the
execution of the interpreter instead of the execution of the application
program. Hence, a resulting trace is not specific to a particular application
but the underlying
interpreter~\cite{bolz+:2009:tracing-meta-level:,bolz+:2013:impact-meta-tracing}.
Therefore, it is not necessary for language implementers to program an
optimized, language-specific \ac{JIT} compiler but rather to provide a
straightforward language-specific interpreter in \RPython, a subset of Python
that allows type inference. \emph{Hints} to the meta-tracing \ac{JIT} enable
fine-tuning of the resulting \ac{JIT} compiler~\cite{bolz_runtime_2011}.
\RPython's tracing JIT also contains a very powerful escape
analysis~\cite{Bolz:2011p1787}, which is an important building block for the
optimization described in this paper. Meta-tracing has been most prominently
applied to Python with PyPy~\cite{rigo+:2006:pypys-approach}.


\section{Optimization Approach}
\label{sec:optim-appr}
\begin{figure*}[tpb]
  \centering
  \includegraphics[%
  width=.95\linewidth%
  ]{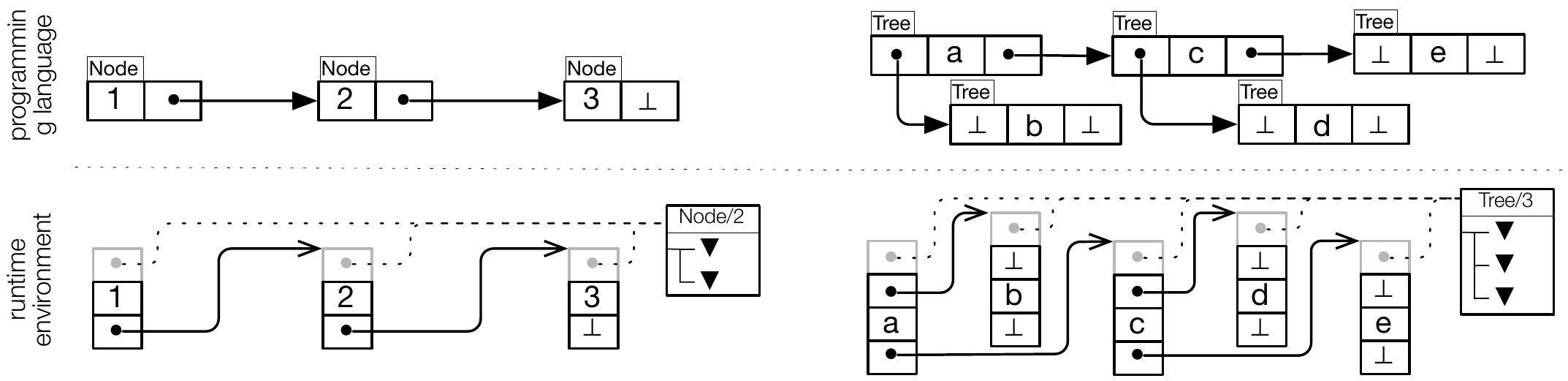}
  \caption[Value class representation]{Straightforward value class representation for a linked
    list and a tree. Top: the language view; bottom: runtime environment view with
    \emph{storage} and \emph{shape}.%
}
  \label{fig:baseline}
\end{figure*}
Our optimization uses an unconventional memory representation for value objects
within the virtual machine to save memory and to speed up access. The
optimization stays invisible to the programmer.

A straightforward representation for a value object in memory is a chunk of
memory that stores a reference to the object's class first, followed by
references for each of its fields. We call the latter the \emph{storage} of the
object. An example of this straightforward representation can be seen in
\autoref{fig:baseline}, which shows a linked list and a tree structure.

The idea of our optimization is to look for common patterns in the object graph
at run-time. If a frequently appearing pattern is identified, we introduce an
abbreviated form to store the pattern. Newly created instances that exhibit
this pattern use the abbreviated form to save memory.

The abbreviated form uses object inlining for instances with these common
patterns. Instead of storing references to a sub-object, the sub-object's
fields are inlined into the referencing object's fields. This saves the pointer
from the outer object to the inlined one, the overhead of maintaining a
separate object and the reference to the inlined object's class. This inlining
is done recursively, if possible. During the inlining process, we need to
maintain certain meta-information to keep track of which fields belong to which
level of an inlined object and in order to remember the classes of the inlined
objects. Therefore, we replace the pointer to the object's class with a pointer
to this meta-information, which we call the \emph{shape} of the object. If no
inlining occurs, we still give the object a shape, which only references the
class and the fact that no inlining is being performed. This is called the
\emph{default shape} of a class.

It is important to not just arbitrarily inline objects
but to do so only for frequent combinations of outer classes and
inner classes. Since the shape needs memory too, introducing shapes that are
solely used by a single object would actually waste memory.

To understand the rest of the system, we now need to look at (a) how structure
patterns are recognized, (b) how the construction of values ensures the proper
usage of shapes, and (c) how the access to of inlined fields is
implemented.

\newcommand*\shapeglyph{{\raisebox{-0.13ex}{\includegraphics[height=.75em]{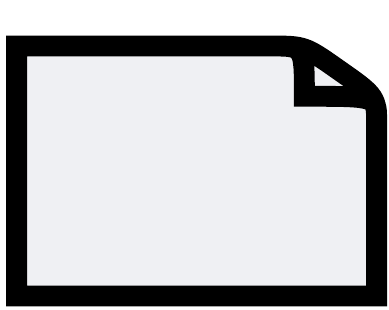}}}\xspace}

\newcommand*\glyphshape[1]{#1{}{\raisebox{-0.66ex}{\includegraphics[height=1.2em]{figures/shapeglyph}}}\xspace}

\begin{figure}[tpb]
  \vspace*{-2\baselineskip}
\hbox to \linewidth{%
  \hspace*{-.3em}
  \includegraphics[width=15em,valign=t]{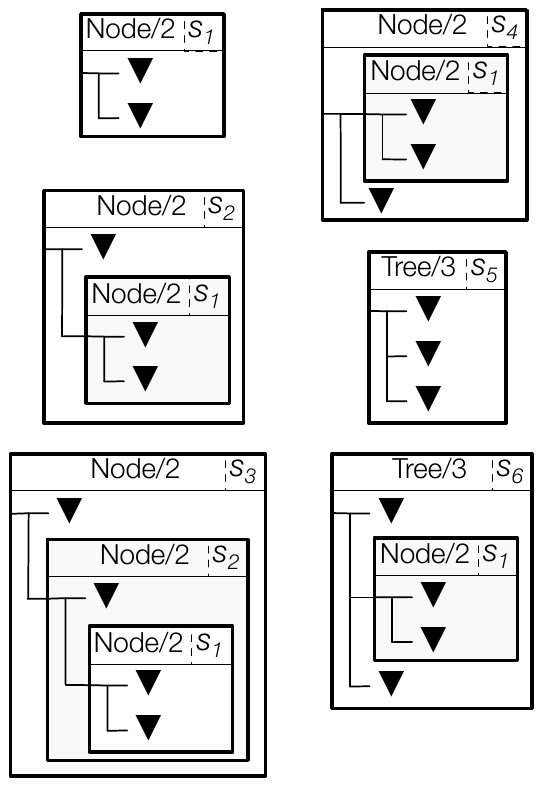}%
\hspace{1.5em}
\begin{minipage}[t]{.4\linewidth} \centering
  {
    \smaller
    \setlength{\tabcolsep}{\dimexpr .5em\relax}
    \newlength\mysep\relax   \setlength{\mysep}{2\tabcolsep}
    \begin{tabular}[t]{@{}cccc@{}}
      \multicolumn{4}{c}{transformation rules}\\
      \toprule
      \phantom{shape} & \phantom{position} & \phantom{sub-shape} & \phantom{substitution} \\[-\normalbaselineskip]
      \multicolumn{4}{c}{%
      \begin{tabular}{@{\hspace*{-\tabcolsep}}c@{\hbox to \mysep{\hss\(\times\)\hss}}c@{\hbox to \mysep{\hss\(\times\)\hss}}c@{\hbox to \mysep{\hss\(\mapsto\)\hss}}c@{\hspace*{-\tabcolsep}}}
        shape & position & sub-shape & substitution
      \end{tabular}}\\
      \midrule
      {\shape[1]{s}} & 1 & {\shape[1]{s}} & {\shape[2]{s}}\\
      {\shape[1]{s}} & 1 & {\shape[2]{s}} & {\shape[3]{s}}\\
      {\shape[1]{s}} & 0 & {\shape[1]{s}} & {\shape[4]{s}}\\
      {\shape[5]{s}} & 1 & {\shape[1]{s}} & {\shape[6]{s}}\\
      \bottomrule
    \end{tabular}
}%
\par
\vspace*{\baselineskip}
{
  \smaller %
    \begin{tabular}[b]{@{}c@{\space}c@{\space}c@{\space}c@{}}
      \multicolumn{4}{c}{history}\\
      \toprule
      {shape} & {position} & {sub-shape} & {occurrences} \\
      \midrule
      {\shape[1]{s}} & 0 & {\shape[1]{s}} & 17\\
      {\shape[1]{s}} & 1 & {\shape[1]{s}} & 28\\
      {\shape[1]{s}} & 1 & {\shape[2]{s}} & 5\\
      {\shape[2]{s}} & 1 & {\shape[1]{s}} & 4\\
      {\shape[2]{s}} & 2 & {\shape[1]{s}} & 19\\
      {\shape[2]{s}} & 2 & {\shape[2]{s}} & 8\\
      {\shape[3]{s}} & 0 & {\shape[1]{s}} & 1\\
      {\shape[3]{s}} & 1 & {\shape[1]{s}} & 3\\
      {\shape[3]{s}} & 1 & {\shape[2]{s}} & 1\\
      {\shape[4]{s}} & 1 & {\shape[1]{s}} & 1\\
      {\shape[5]{s}} & 1 & {\shape[1]{s}} & 13\\
      &&&\ldots \\
      \bottomrule
    \end{tabular}
}
\end{minipage}
\hfil
\includegraphics[width=12em,valign=t]{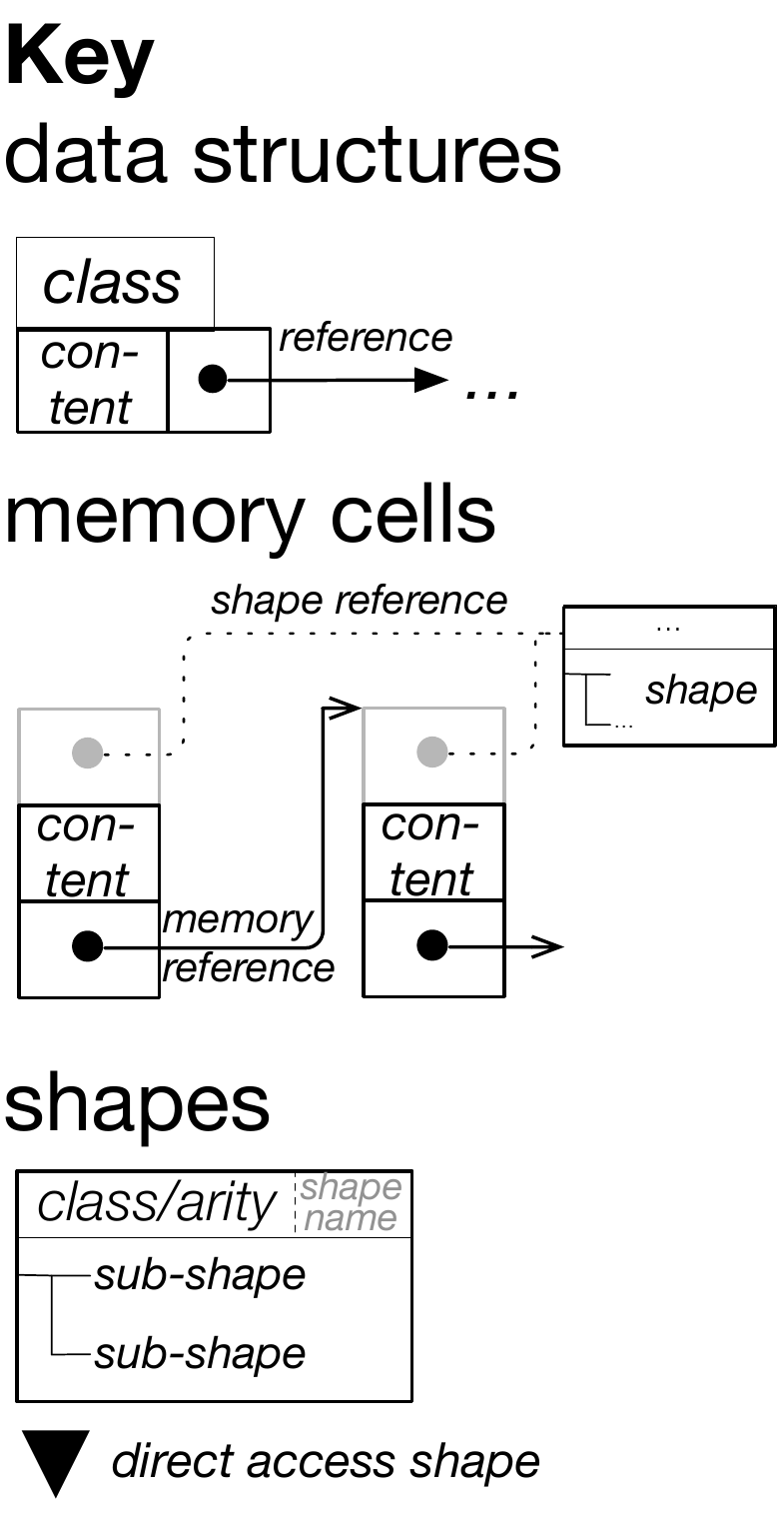}%
  \hspace*{-1em}
}
\caption{
Left: Shapes comprise a class reference, an arity, and a structure of
 sub-shapes.
Center: ``Transformation rules'' describe substitutions for shapes which are
consulted during the inlining process; ``history'' contains a histogram of all sub-shapes
encountered at a certain position in a certain shape collected during all
value object creation.
Right: Key to the visual language used.%
 }
  \label{fig:shapes}
\end{figure}%

\subsection{Shapes and their recognition}
\label{sec:shape-recognition}
A \emph{shape}
describes the abstract, structural representation of value objects. It is
shared between all identically structured instances of the same value
class\footnote{We refer to a value class by its name and the arity of its type
  in a Prolog style, \eg \constr{Node/2} for binary node objects.} and captures
the structure of these instances. Value objects have a permanent reference to
their shape during their life time.

Shapes can be nested; they consist of sub-shapes for each field in a value
object's storage. A special, flat shape denotes unaltered access to object
fields (\emph{direct access shape}, \(\blacktriangledown\)
in all figures) and termination of shape nesting. It conveys no more
information than that a field exists and may contain data. Value objects with
these shapes are treated as black boxes, \eg scalar data or unoptimized objects
that are stored directly. This is depicted in the bottom part of
\autoref{fig:baseline}; all three nodes in the list share the same shape, which
denotes that each node consists of two references with \emph{direct access}
shapes. The same holds for the nodes of the tree in that figure, but with three
references.

As long as no optimization has taken place, a value object refers to the
\emph{default shape} of its value class that solely consist of \emph{direct
  access} sub-shapes. The shapes in \autoref{fig:baseline} are the default
shapes for their value classes. Initially, all value object use a default
shape. To reach a state where more complex shapes can be used, our approach
depends on auxiliary data.

To guide the overall optimization process, we keep track of all shapes that we
encounter during object creation. That way, we create a histogram of all shapes
used in the fields of value objects. We explain this profiling data, which we
call the \emph{history}, in \autoref{sec:history}.

Based on the history profiles, we determine the fields in a value class where
inlining value objects could be worthwhile. We infer new shapes for value
objects with certain referenced value objects inlined, and record a transition
from the old to the new shape. We call this process \emph{shape recognition}
and explain it in \autoref{sec:shape-recognition-1}.

We collect all results from the shape recognition in a table that we call the
\emph{transformation rules}. We explain its structure briefly in
\autoref{sec:history-1}.

\subsubsection{History}
\label{sec:history}

The history is a table that counts how often certain sub-shapes are found in
the fields of new value objects. It is essentially a histogram of all
sub-shapes. It is rather simple to maintain, as due to the immutability of
value objects, modifications of this table are only necessary during value
object creation. At this point, all objects that will constitute a new value
object are available and we can count the occurrences of \emph{sub-shapes} at
specific positions in the value object.

As example, the history table in~\autoref{fig:shapes} shows that for shape
\shape[1]{s} at position 1, the shape \shape[1]{s} itself has been encountered
17 times as sub-shape, while shape \shape[2]{s} has been encountered 5 times as
sub-shape in that position.

The most important operation on the history table is updating the count of a
shape\(\times\)position\(\times\)sub-shape--entry,
besides initializing it to 1 on the first encounter. It is possible to remove a
history entry after it had been used for creating a transformation rule, if
desired.

\subsubsection{Shape recognition}
\label{sec:shape-recognition-1}
During the creation of a value object we first update the shape history table
and then check the counters associated with the shapes of the object's fields.
Whenever one of these counters exceeds a preset threshold, create a new shape
that combines the value object's current shape with the sub-shape that exceeded
the threshold. In this new shape, we replace the \emph{direct access} sub-shape
at the position where the threshold was reached with the sub-shape found in the
history entry. We then create a new transformation rule that maps from the old
shape, the position, and the sub-shape at that position to the newly created
shape.

Considering \autoref{fig:shapes} as example, shape \shape[2]{s} would be the
result of turning the history entry (\shape[1]{s}, 1, \shape[1]{s}, 17) into
the transformation rule (\shape[1]{s}, 1,\shape[1]{s})\(\mapsto\)\shape[2]{s}.

\subsubsection{Transformation rules}
\label{sec:history-1}

We maintain the set of all transformation rules as a lookup table that is
used during value object creation. This table is only ever updated during shape
recognition and typically, rules are never removed from it. However, it is
usually much smaller than the history table. Find an example transformation
rule table in the top center of \autoref{fig:shapes}.

\medskip
Note that we consider both history and transformation rules to
\emph{conceptually} be tables. Depending on circumstances it may be advisable
to merge them into one table or split them by the first column's entries and
attach them directly to those shape.

\newcommand*\valueobject[1]{``\(#1\)''\xspace}

\subsection{Compaction through inlining}
\label{sec:compr-through-inlin}
\begin{figure*}[tbp]
  \centering
  \includegraphics[width=\linewidth]{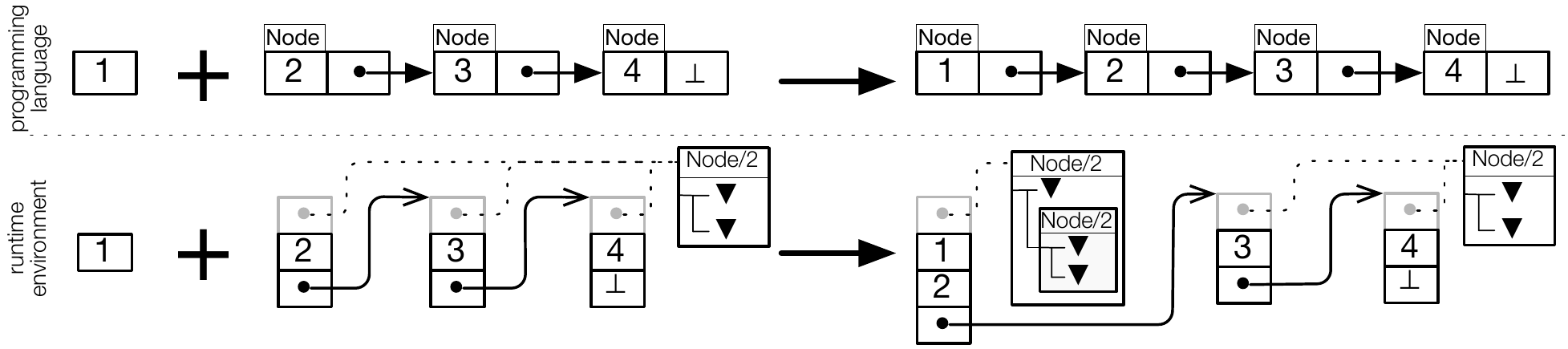}
  \caption{When creating a new node value object that should contain \valueobject{1} and
    the list
    \valueobject{\constr{Node/2}[2, \constr{Node/2}[3, \constr{Node/2}[4, \bot]]]},
    a new value object that merges the ``1'' with the ``2'' object and a
    different shape is created instead.}
  \label{fig:transformation}
\end{figure*}
The information of what shapes occur often and which shape transformations to
use can be applied at run-time to create value objects in a compacted
representation. The process of creating such a compacted value object is
outlined in the following. As running example, we will use the combination of
the primitive datum \valueobject{1} with a linked list into a new linked list as
depicted in \autoref{fig:transformation} and using the shapes and
transformation rules as given in \autoref{fig:shapes}.

First, it is only necessary to consider compaction when creating new value
objects. Since they are immutable, there is no need to consider compaction on
mutation. Therefore, the inlining process starts with the following two
components:
\begin{enumerate}
\item the value class of the object that is to be created, and
\item the elements that should constitute said object's new fields.
\end{enumerate}
In our example, the class is \constr{Node/2} and the new fields are
\valueobject{1} and a \constr{Node/2} value object
(\valueobject{\constr{Node/2}[2, \ldots]}). As pointed out earlier, every value
class has an associated default shape equivalent to a straightforward
representation. In the case of the class \constr{Node/2}, this default shape
corresponds to shape \shape[1]{s} in \autoref{fig:shapes}.
\begin{algorithm}
  \centering
  \caption{Determining shape and fields of a value object during its creation.
    The shape is derived based on transformation rules and the fields are
    inlined based on the resulting shape}
  \label{alg:merge}
  \begin{minipage}[t]{2em}\raggedleft\color{LightGrey}
   \vspace\baselineskip
1\\2\\3\\4\\5\\6\\7\\8\\9\\10\\11\\12\\13\\14\\15
  \end{minipage}\hspace{.5em}
  \begin{minipage}[t]{.4\linewidth}
   \vspace\baselineskip
  \obeylines
  \textbf{Input:} \(s: \id{Shape}, f: [\id{Value Object}]\)
  \( i \gets 0 \)
  \textbf{while} \( i < |f|\) \textbf{do} {
    \leftskip\dimexpr\leftskip + 2em\relax%
    \(s_i \gets \prop{\id{\ensuremath{f_{i}}}}{\id{shape}} \)
    \( s' \gets \id{transformations}_{s,i,s_i}\) or \( s \)
    \textbf{if} \( s' \ne s \) {
      \leftskip\dimexpr\leftskip + 2em\relax%
      \(f \gets \left[ f_{0,\ldots,i-1}, \prop{\id{\ensuremath{f_{i}}}}{\id{storage}}, f_{i+1,\ldots,|f|} \right]\)
      \(s \gets s' \)
      \textit{\color{gray}{// restart with new storage}}
      \(i \gets 0 \)
    } \textbf{else} {
      \leftskip\dimexpr\leftskip + 2em\relax%
      \( i \gets i + 1\)
    } \textbf{end}
  } \textbf{end}
  \textbf{return} {\(s, f\)}
\end{minipage}
\end{algorithm}
With the default shape and the fields, the inlining algorithm as specified in
\autoref{alg:merge} can now commence. In our example, the initial shape \(s\)
provided as input to the algorithm is the default shape \shape[1]{s} and the
fields \(f\) are \valueobject{1} and \valueobject{\constr{Node/2}[2,\ldots]}.

We now iterate over the fields (line 3) and consider each new field \(f_{i}\)
separately. For that, we look at the sub-shape \(s_{i}\)
of the new field \(f_{i}\)and
try to look up a substitute shape \(s'\)
(line 5). If we have no substitution, \eg because none has been recorded yet or
the new field \(f_{i}\)
is primitive data, the shape is not substituted and we continue with the next
element. However, if we find a substitute (line 6), we replace the value object
\(f_{i}\)
with a copy of its \emph{storage} in the new fields \(f\)
(line 7); the value object \(f_i\)
is now \emph{inlined}. The new shape \(s'\)
becomes the new value object's shape \(s\)
(line 8) and the inlining process is \emph{restarted} (line 10) with the new
shape and fields. This allows possible other transformation rules to be applied
due to the shape change.

Once no further transitions are found, the value object's shape \(s\)
and the current fields \(f\)
are returned as the shape and storage of the new value object (line 16).

For our example, the following happens: while iterating over the new fields
\(f\), we encounter \valueobject{1} as the first field \(f_{0}\). Since this is
a primitive datum, no new shape can be found and no shape change happens. The
next new field \(f_{2}\) to consider is
\valueobject{\constr{Node/2}[2,\ldots]}. The sub-shape \(s_{1}\) of this value
object is \shape[1]{s} and we can now look up a transformation rule for
(\shape[1]{s}, 1, \shape[1]{s}) and find a substitution \(s'\), \shape[2]{s}
(line 5). Thus, we inline the storage of \(f_{1}\) by copying it into the new
fields \(f\) at position \(1\).\footnote{The original value object
  \valueobject{\constr{Node/2}[2,\ldots]} remains untouched and can still be
  referenced from other objects.} The fields of \(f\) are now
\valueobject{1}, \valueobject{2}, and \valueobject{\constr{Node/2}[3,\ldots]}.
Furthermore, we change the shape of the new value object to \shape[2]{s} (line
8). At that point, we restart the  inlining process by resetting the counter
(line 9). This means, we again encounter \valueobject{1} as first field
\(f_{0}\) and no substitution happens. Moreover, the second field \(f_{1}\) is
now \valueobject{2}, so no substitution happens either. We continue with the
third field \(f_{2}\), which is \valueobject{\constr{Node/2}[3,\ldots]}. The
sub-shape of this value object is \shape[1]{s}, and since \(s\) is
\shape[2]{s}, we can look up a transformation rule for (\shape[2]{s}, 2,
\shape[1]{2}) in the table. However, no such transformation rule exists and,
hence, no further inlining is possible. Since we visited all fields, the
algorithm terminates and returns the value object's new shape \shape[2]{s} and
its new fields \([1, 2, \constr{Node/2}[3,\ldots]]\).

During the inlining process, potentially short-lived objects might be created.
This can happen when the storage of a value object is inlined into its
surrounding list of fields. Typically, a new list of correct lengths is created
and the old list will be un-referenced. In subsequent inlining steps, this new
list itself may be short-lived. To retain simplicity in our approach, we
refrained from introducing sophisticated mechanisms to avoid the allocation but
rather rely on modern \ac{JIT} compilers. We expect those allocations to happen
in tight loops, but more importantly, in a very restricted scope. Hence,
\ac{JIT} compilers with good escape analysis and allocation removal, such as
meta-tracing \ac{JIT} compilers~\cite{bolz+:2011:allocation-removal}, should be
able to completely remove all allocations during the inlining process.

This shape inlining technique has two main advantages. First and foremost,
inlined value objects take up less space than individual, inter-referenced
value objects. But even more, the shape of a value object provides structural
information in a manner the meta-tracing \ac{JIT} compiler can speculate on.
This is crucial to optimize field accesses in a value object.

\subsection{Implementing field access}
\label{sec:impl-field-access}
\begin{figure*}
  \centering
  \includegraphics[width=\linewidth]{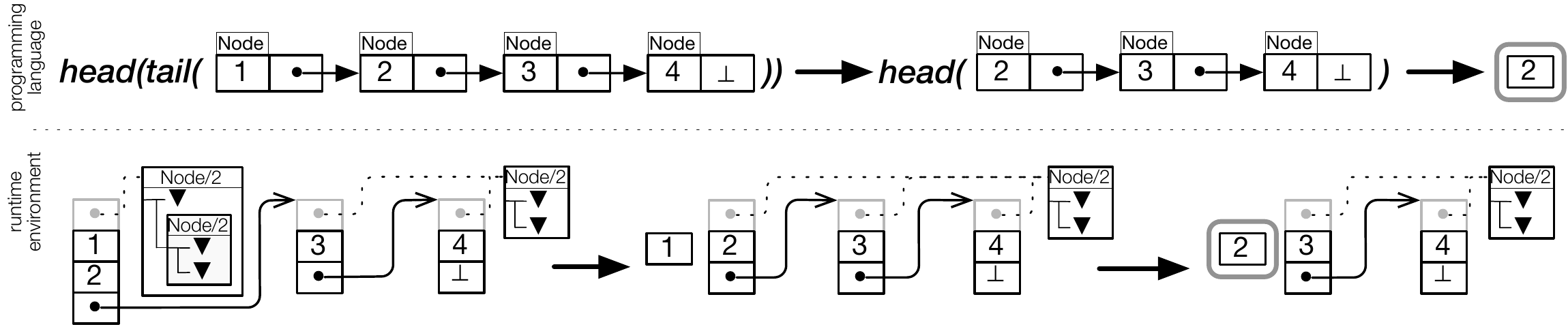}
  \caption[Referenced value object reification]{Referenced value object
    reification. Accessing the second item \(2\) of the list
    \(l \gets \constr{Node/2}[1, \constr{Node/2}[2, \constr{Node/2}[3, \constr{Node/2}[4, \bot]]]]\) by two operations \(\id{head}(\id{tail}(l))\)
    results in two reified rest lists to be created.}
  \label{fig:fieldaccess}
\end{figure*}

While optimization of data structures takes place during construction, we have
to apply the reverse during deconstruction, \ie when accessing a value object
referenced by another. This is no longer trivial, as several (formerly
referenced) value objects may have been inlined into their referencing value
objects. Therefore, we construct new value objects whenever a reference is
navigated, essentially reifying it. We use the information a value object's
shape provides to identify which parts of the value object's storage comprise
the value object to be reified. The structural information allows a direct
mapping from the language view of the data structure to the actually stored
elements. In \autoref{fig:fieldaccess}, the structural information in the shape
of the leftmost list allow the reasoning that the first element of the storage
is equivalent to the head of the language level node value object and the
remaining three storage elements are equivalent to the tail of that value
object, as recored in the shape. Hence the middle view in that figure; both the
element ``1'' and the rest list have been reified. The same goes for the
rightmost view.

Note that this reification is completely invisible to programmers. Taking, \eg
the tail of a node value object or accessing the third element of a ternary
tree repeatedly, the operations remain the same on the language level, no
matter what is the shape inlining status of the value objects on the
implementation level.

\subsection{Benefits}
\label{sec:benefits}

With the shape inlining approach, fewer value objects need to be created for long
living data structures, since the references to the now-inlined value objects
are elided. Combining this with the reification and the shape recognition, more
memory is saved the longer a program runs; the shapes will be tailored to fit
the specific application running. That said, there may be cases where no memory
can be saved, especially in programs that only work on primitive data,
flat data structures, or with a high amount of sharing between data
structures.


\section{Implementation in RPython with a tracing JIT compiler}
\label{sec:impl}

We present two implementations of our approach, both integrating the tracing
\ac{JIT} compiler of RPython as presented in \autoref{sec:tracing-jit}.

\subsection{JIT Interaction}
\label{sec:jit-inter}

While the techniques we described so far can lead to a good amount of memory usage reduction,
shape recognition, shape inlining, and reified reference access combined, do not
yield a performance increase on their own. In fact, implementing the approach
naively yields significantly worse performance%
, due to the constant check of the transformation rules every time a new value
object is created. Additionally, reading inlined fields of compacted value
objects results in the allocation of intermediate data structures. This is of
course not the case in the naive representation. Hence, the presence of the
\ac{JIT} compiler is necessary to begin with.

To improve performance, the \ac{JIT} compiler needs to reduce the overhead of
these operations. The first step is to treat the transformation tables as
constant when a function is compiled. This allows the \ac{JIT} compiler to
compile value object creation down to a series of type checks for the types of
the referenced value objects. We instruct the \ac{JIT} compiler to treat
transformation tables as constant after filling it with enough information.

Second, we have to avoid the otherwise necessary reification of referenced
value objects when it is being read from a value object it has been inlined
into. For that, the observation that most of these intermediate value objects
are actually short-lived is crucial; most value object are created just to be
either immediately discarded or consumed in another, typically larger data
structure. As a concrete example, typical linked list operations deconstruct
the list they are working on. Hence, if the tail is read off a linked list node
which has the tail inlined (as the transition from left to middle in
\autoref{fig:fieldaccess}) and needs to be reified, that tail is usually soon
deconstructed itself into its head and tail components (as the transition from
middle to right in \hyperref[fig:fieldaccess]{the same figure}). This allows
the tracing \ac{JIT} compiler to optimize the reading of fields that need
reification. Since the value objects allocated when reifying a field are
short-lived, the built-in escape analysis \cite{bolz+:2011:allocation-removal}
will fully remove their allocation and thus remove the overhead of reification.

\subsection{Best-case Prototype}
\label{sec:best-case-prototype}
To assess best-case performance, we implemented our optimization approach using
a simple execution model prototype\footnote{available at
  \url{https://bitbucket.org/krono/lamb}}. It provides a \(\lambda\)-calculus
with pattern matching as the sole control structure and is implemented as a
direct application of the
\textsc{cek}-machine~\cite{felleisen+:1987:control-operators}. The only
structured data types available are value classes. We used the \RPython tool
chain to incorporate its meta-tracing \ac{JIT}
compiler~\cite{bolz:2012:meta-tracing-just-in-time}. The implementation has
been carefully unit-tested during development to make sure that various
complex substitutions and compactions work correctly.

\subsection{Structures in Racket and Pycket}
\label{sec:pycket}

Since the best-case prototype is arguably unfit for comparison with existing
languages and their implementations, we applied our optimization to an
implementation of the Racket language~\cite{plt-tr1}, a dynamically typed,
multi-paradigm programming language in the Scheme family. Racket supports,
among others, immutable-by-default lists, a
design-by-contract~\cite{mitchell+:2001:design-contract} implementation, and
heterogeneous \emph{structure} datatypes.

The structure types are of special interest because, if applied carefully, they
can be used like value classes.\footnote{Structures in the Racket language
  actually do not by default compare based on their value and do have identity
  which is relied upon. However, value-based comparison can be enabled
  explicitly. Also, plans exist to provide a structure derivative that has a
  concept of identity compatible with value classes. [Sam Tobin-Hochstadt,
  personal communication]} Moreover, structures can form hierarchies and\---
by default\--- are immutable with the option to make some or all fields
mutable. Racket structures go beyond other structured heterogeneous datatypes;
they support the notion of structure type properties that can influence the way
structures interact with the system. For example, a special structure type
property can make structure instances \emph{callable}, so they can act
like a procedure.

\emph{Pycket}~\cite{spenser-bauman+:2015:pycket:-tracing} is an implementation
of Racket using the RPython toolchain and its tracing \ac{JIT}. While not
feature-complete, it provides a fair amount of functionality and can compete
with the reference implementation performance-wise, in certain areas even
outperforming high-performance \ac{AOT} Scheme compilers. The support for
Racket structures in Pycket is recent~\cite{pape+:2016:optimizing-record} and
showed potential for the optimization presented here. Furthermore, the
implementation technique (\textsc{cek} machine) and environment (RPython,
tracing \ac{JIT}) come close to the prototype and suggest a good base for
comparison.

Our approach is present in a modified Pycket
implementation.\footnote{\url{https://github.com/samth/pycket/releases/tag/shapes-scp}
  (last accessed 2015-12-15)} The existing structure
implementation~\cite{pape+:2016:optimizing-record} already tries to optimize
memory consumption and execution time. It already deals with the distinction of
smaller and larger structure instances; for the former, objects with a known,
small number of fields are used, for the latter, separate storage objects are
created. Hence, an abstraction for field accesses already existed. We were able to
take the implementation of the prototype with little modification and use it as
storage for all structure kinds. Only few adaptions were necessary: we added
the management logic for shapes and re-routed access to fields through them.
All in all, the changes amounted to less than 550 lines of code added and a
handful of lines of codes removed.

\subsection{Configuration Parameters}
\label{sec:conf-param}

Our approach makes use of three parameters that may influence performance:

\paragraph{Maximum object size} Only value objects up to this size are
considered for inlining. Setting this to zero disables our optimization, setting
it to a very high number might result in very large value object at runtime,
which might be undesirable.

\paragraph{Maximum shape depth} The number nested shape occurrences per
value object is bounded by this parameter. Setting this to a low value may not
catch all optimizable object shapes, setting it to a very high number may lead
to an excessive number of shapes at runtime should there be a lot of value
objects with no fields at all.

\paragraph{Substitution threshold} The threshold for transformation rule
creation (as in section~\ref{sec:history-1}), when set to zero or a very low
value can lead to excessive transformation rule creation for value object
combinations that are only rarely used. A very high number might inhibit the
creation of such rules at all and practically disables our optimization.


\section{Results}
\label{sec:results}

\def\idBox#1#2{%
\setlength{\fboxsep}{2pt}%
\raisebox{.5pt}{\colorbox[HTML]{#1}{\textcolor[gray]{0.9}{\rule[0.3pt]{0pt}{5pt}#2}}}%
\xspace}

We present two kinds of results. First, we show that the shape recognition part
(\cf{\autoref{sec:shape-recognition}}) of our approach is feasible and can be
used instead of manually specifying shape transformation rules. And second, we
present the execution time and memory consumption for selected micro-benchmarks
on our two implementations and three more language implementations.

\subsection*{Setup}
\paragraph{Hardware} The processor used was an Intel Xeon E5410 (Harpertown)
clocked at \SI{2.33}{\GHz} with \(2 \times\)\SI{6}{\mega\byte} cache;
\SI{16}{\giga\byte} of RAM were available. All runs are un-parallelized, hence
the number of cores (four) was irrelevant to the experiment. Although
virtualized on Xen, the machine was dedicated to the benchmarks.

\paragraph{Software} The machine ran Ubuntu 14.04.3 LTS with a \SI{64}{\bit}
Linux 3.13.0. \emph{ReBench}\footnote{ReBench is a benchmarking framework.
  \url{https://github.com/smarr/ReBench}} was used to carry out all execution
of the benchmarks and collection of measurements. %
\RPython as of revision 0c8d6f715aac served for translation of our prototype
(tag \texttt{shapes-scp}) and the optimized Pycket (tag \texttt{shapes-scp}).

\paragraph{Optimization Configuration}
During the measurements of our implementations, we used the following settings
for the configuration parameters as described in \autoref{sec:conf-param}:
\begin{description}
\item[Maximum object size] We used a maximum size of 7 fields.
\item[Maximum shape depth] We used a maximum depth of 7 shapes.
\item[Substitution threshold] We used a threshold of 17 shape occurrences.
\end{description}

\subsection{Shape recognition fitness}
\label{sec:shape-recogn-fitn}

\begin{figure}
  \centering
  \includegraphics[width=\linewidth]{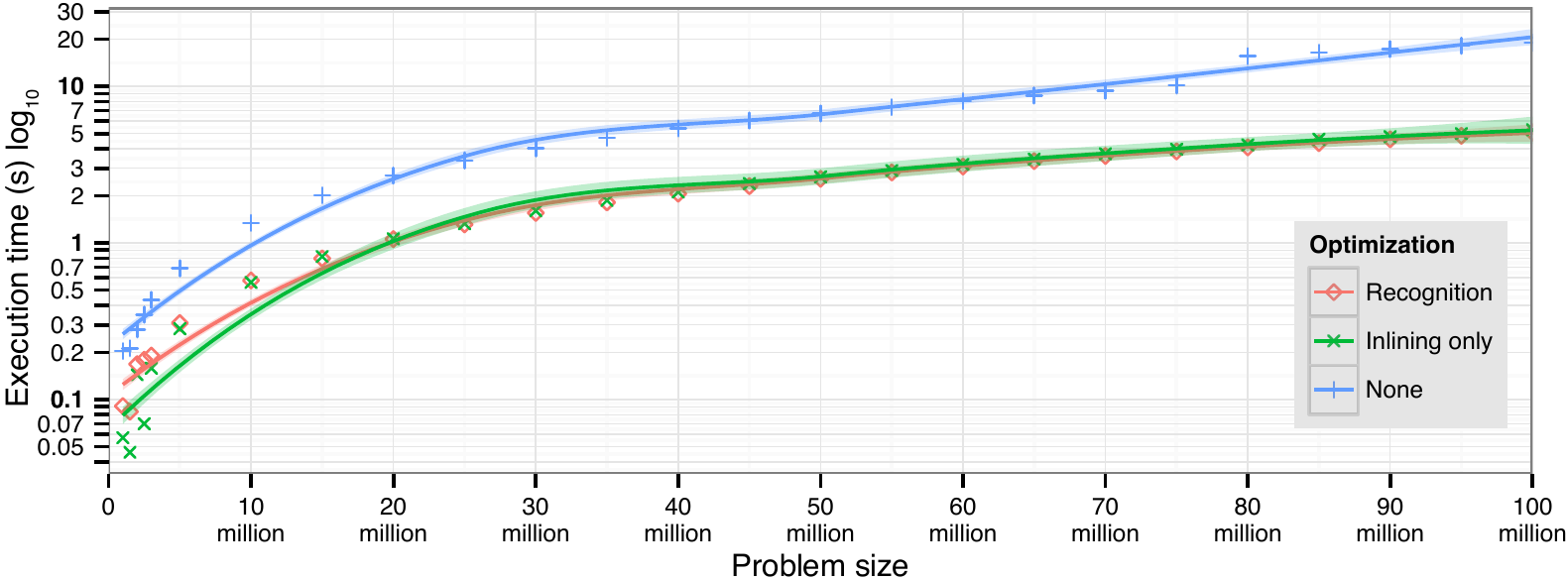}
  \caption{Runtime results for reversing list of different lengths. \emph{None}
    is without our optimization approach. \emph{Inlining only} uses our
    optimization approach with ahead-of-time, manually specified transformation
    rules without using shape recognition. \emph{Recognition} uses our
    optimization approach with transformation rules derived using shape recognition.
    (The data points were smoothed using local
    regression~\cite{cleveland+:1992:local-regression}; the semi-transparent
    areas are based on standard deviation of each data point. Note the
    logarithmic scale on the ``Execution time'' axis.)%
  }
  \label{fig:revplot}
\end{figure}

To assess whether our recognition approach is favorable to manually specifying
shape transformation rules, we ran several list operations on increasingly
longer, large lists in our prototypical implementation in three configurations:
no optimization at all (None), optimization using our approach but only using
ahead-of-time, manually specified transformation rules without using shape
recognition or history data (Inlining only), and optimization with
transformation rules derived using shape recognition and history data
(Recognition).
We provide the execution time results for reversing a long list
in \autoref{fig:revplot}. In this case, we found that
\begin{enumerate}[a)]
\item both optimized versions are always significantly faster than the not
  optimized version,
\item initially, the version with manually specified transformation rules is
  faster than the version with shape recognition, but
\item for most data points, the version with shape recognition and
  transformation rule inference is as fast as or even faster than the version
  with manually specified transformation rules.
\end{enumerate}
The results for other list operations (appending, mapping, filtering) were very
similar and have hence been omitted.

The results suggest that the shape recognition approach could be fitting in the
context of our optimization and could be favorable to specifying transformation
rules manually.

\subsection{Comparative Micro-benchmarks}
\label{sec:comp-micro-benchm}

We report the performance of five micro-benchmarks with their execution time
and peak memory consumption.

\paragraph{Compared Implementations}
For the benchmarks, we included an unmodified
Pycket~\idBox{4DAF4A}{\(\square\)},
Racket~\idBox{984EA3}{\(+\)}, and PyPy~\idBox{FF7F00}{\(\boxtimes\)}\footnote{%
  Pycket revision 291d80fbd43a; Racket version 6.3; PyPy version 4.0.1} in the
comparison. For all these, value classes or equivalent means supporting
immutable data are available. The unmodified Pycket is the baseline of our
implementation and does not include our optimization. Racket's \emph{cons}
cells, \emph{structs} and classes can act as value classes. Racket acts as a
virtual machine with a handwritten \ac{JIT} compiler. PyPy is the \RPython
implementation of Python and has a meta-tracing \ac{JIT} compiler. While Python
has no actual concept of value classes, we used regular classes without
mutating them. PyPy detects this case and is able to apply special
optimizations, effectively treating them like value classes. We intended
to also include the standard Python (CPython) but it was too slow and would
have rendered the comparison meaningless.

\paragraph{Methodology} Every benchmark was run ten times uninterrupted at
highest priority, in a new process. The execution time (\emph{total time}) was
measured \emph{in-system} and, hence, does not include start-up; however,
warm-up was not separated, so \ac{JIT} compiler execution time is included in
the numbers. The maximal memory consumption (\emph{resident set size}) was
measured \emph{out-of-system} and may hence include set-up costs. We report the
arithmetic mean of the ten runs; for the execution time we include
confidence intervals showing the \SI{95}{\percent} confidence level.
The memory measurements only indicate a negligible error\footnote{except for
  Racket, which we attribute to its garbage collector (\cf{the table in the
  appendix})} that was hence omitted. We provide all numbers in tables for
\hyperref[tab:time]{execution time} and \hyperref[tab:mem]{memory consumption}
in the appendix. Our benchmarking code and infrastructure are publicly
available.\footnote{\url{https://bitbucket.org/krono/lamb-bench}}

\subsubsection{Benchmarks}

The benchmarks chosen are \emph{append}, \emph{filter}, \emph{map}, and
\emph{reverse} on very long linked lists and the creation and complete prefix
traversal of a binary \emph{tree}. Due to the limited feature scope of our
best-case prototype, more sophisticated applications are currently not
available for benchmarking. For our optimization of Pycket, the structure
benchmarks shipped with Racket would be interesting for our measurements. %
However, the structure benchmarks do not run yet on Pycket due to missing (not
structure related) features \cite{pape+:2016:optimizing-record}. %

\subsubsection{Non-regression}

Our optimization should not influence anything except value classes. To ensure
this for Pycket,, we ran the \emph{shootout} benchmarks described in the
original paper on Pycket~\cite{spenser-bauman+:2015:pycket:-tracing}. These
benchmarks hardly make use of structures. On average, the execution time for
these benchmarks deviates less than \SI{6}{\percent} (both faster or slower) %
from the original implementation. This low deviation shows that our approach
has very little overhead when structs are not used.

\subsubsection{Performance Results}
\begin{figure*}[tbp]
  \hbox to \linewidth{\hss\vbox{%
  \hbox{\includegraphics[width=1.03\linewidth]{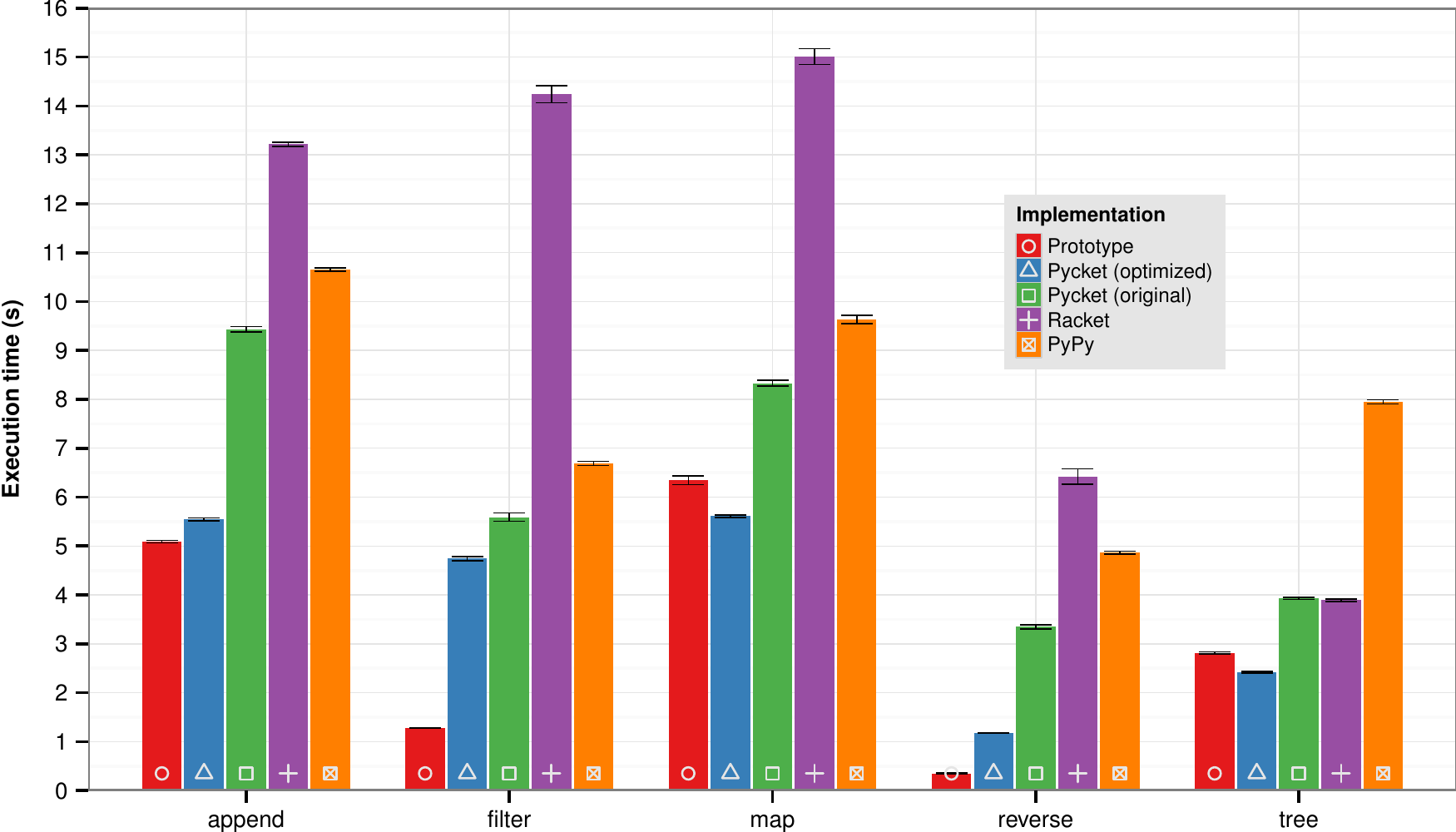}}%
  \hbox{\includegraphics[width=1.03\linewidth]{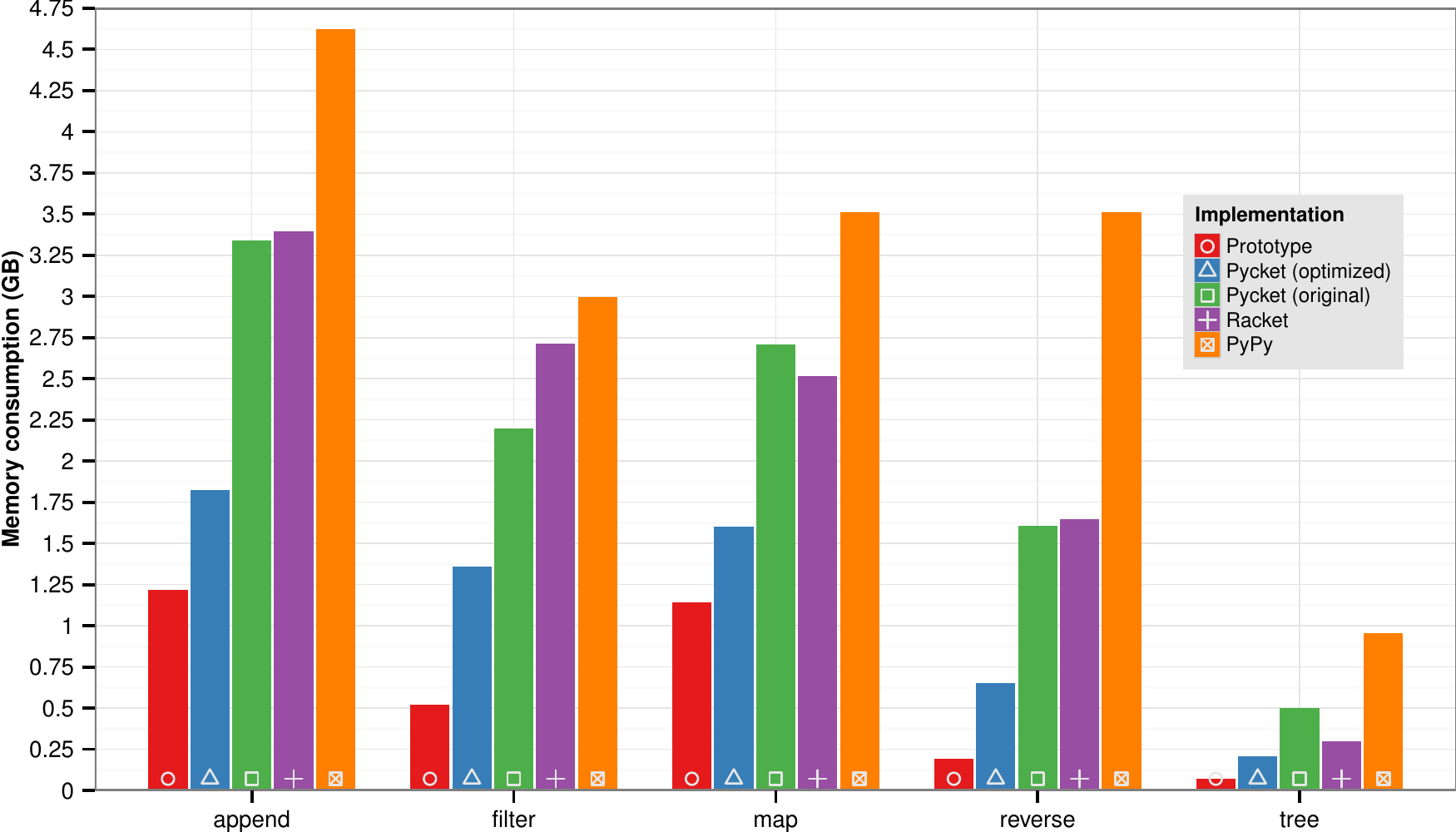}}%
}
}
 \caption{Benchmarking results. Each bar shows the arithmetic mean of ten runs
  for execution time (top) and memory consumption (bottom). Lower is better.%
}
  \label{fig:results}
\end{figure*}

In the top part of \autoref{fig:results}, the execution time of all benchmarks
is reported. Our first implementation, labeled
\emph{prototype}~\idBox{E41A1C}{\(\circ\)},
is significantly faster---from two to ten times faster.
Our second implementation, labeled \emph{optimized
  Pycket}~\idBox{377EB8}{\(\bigtriangleup\)}, performs as expected.
It is not as fast as the best-case prototype, as the language semantics of
Racket have to be maintained as much as possible. However the speed-up compared
with the unmodified, unoptimized version of Pycket is apparent. The optimized
version is \numrange{1.2}{2.9} times faster than the unoptimized version. In
the case of \emph{map} and \emph{filter}, the optimized Pycket version is even
faster than the prototype. We attribute this to the more mature status of
Pycket compared with the prototype, which is a pretty direct implementation of the
\(\lambda\)-calculus.

For memory consumption, shown in the bottom part of
\autoref{fig:results}, our implementations always use significantly less memory
than the other implementations. The optimized Pycket implementation is always
second to our best-case prototype and in the best case uses only
\SI{40}{\percent} of the memory the unoptimized Pycket uses. The memory
consumption of our best-case prototype is very low, as its execution model is
quite restricted, and the only data structure types available are value
classes, the subject of our optimization. On the other hand, the other language
implementation face more complex execution models with more meta-data and other
kinds of data structures besides value classes. Under this assumption, we think
the differences between the optimized Pycket and the unoptimized Pycket are the
most significant result from the memory analysis.

One key reason for our implementations' performance is the interaction between
escape analysis and the compacted storage. The benchmarks exhibit a certain
usage pattern, in particular, the access to a list element is typically
followed by inserting this element into a new list, with possibly processing
it. The tracing \ac{JIT} compiler and its escape analysis can infer that no
reification of the actual value object is necessary and, furthermore, that a
certain number of such operations occur consecutively. Hence, operations can
happen block wise, \eg for a list inlined \(n\)
levels deep, reverse can operate in chunks
of \(n\) items. Proper tail recursion amplifies this effect.

Given our parameters (maximum object size of 7 and maximum shape depth of 7),
we expect the inlining for to result in chunks of 6 consecutive list elements.
This means that (a) five class references and five next-element references can
be saved per chunk, \ie more than \SI{50}{\percent}, and (b) the list
operations can work on these chunks consecutively, comparable to what list
unrolling achieves. Moreover, the tracing \ac{JIT} compiler can make
assumptions on these chunk and remove almost all type checks, reduce the number
of allocations to a minimum, has to follow less references, and reduce the
overall number of operations the tracing \ac{JIT} processes by up to
\SI{60}{\percent}.

\section{Related Work}
\label{sec:related-work}

Data structure optimization is well documented in literature and industry. We
want to put our approach to value class optimization into this context.

\paragraph{Algebraic Data Types}
From a data structure optimization point of view, value classes are similar to
\emph{algebraic data types} as found in languages in the ML
family~\cite{milner+:1997:definition-standard,damas+:1982:principal-type-schemes}.
Hence, optimizations done to this category of data structures are relevant to
value classes, too~\cite{lee+:1996:optimizing-with,Leroy-ZINC}.

\paragraph{Object Inlining}
Wimmer has proposed object inlining~\cite{wimmer:2008:automatic-object} as a
general data structure optimization for structured objects in Java. This
approach shares many similarities with ours: it also inlines objects into their
referring objects, saving space and pointer indirections. It has a number of
advantages over our approach: the approach guarantees to never need more memory
than without the optimization. Also, it does not need any complex run-time
support, since it relies on a static, global analysis to identify classes for
which the inlining is possible. This latter property is however also a
weakness: it restricts the approach to statically typed programs where global
analysis is possible, which hampers the use in dynamic languages and in
settings where reflection or class loading is used. Additionally, the inlining
decision is done per class, while in our approach different shapes and thus
inlining patterns can be created for a single value class.

\paragraph*{Language-level optimization}
\label{sec:lang-opt}

Improving data structures to gain execution speed has been proposed for
operations on linked lists in functional languages, \eg by
unrolling~\cite{shao+:1994:unrolling-lists}. Typically, those optimizations are
restricted to linked lists of cons-cells.

One of the key effects in our optimization is avoiding to allocate intermediate
data structures. In that respect, \emph{hash consing}%
~\cite{ershov:1958:programming-arithmetic,%
  goto:1974:monocopy-associative,filliatre+:2006:type-safe-modular}, as used in
functional languages for a long time, is related to this work. However, hash
consing typically works at the language level using libraries, coding
conventions, or source-to-source transformations. It is not adaptable at
run-time.

\paragraph*{Ahead-of-time optimization}
\label{sec:aot-opt}

Deforestation~\cite{wadler:1990:deforestation:-transforming,%
  gill+:1993:short-deforestation,takano+:1995:shortcut-deforestation} has the
aim to eliminate intermediate data structures and is in this respect related to
our approach. However, deforestation deliberately works through program
transformation and does not incorporate dynamic usage information. It is
typically only available to statically typed functional languages, such as ML.

\paragraph*{Just-in-time compilers}
\label{sec:jit-comp}

Compiling to native code at run-time, \ie \ac{JIT} compilation, is a prevalent
and extensively studied technique, found in several different, but chiefly
object-oriented, dynamically-typed languages~\cite{aycock:2003:brief-history}.
Prominent examples include the Smalltalk-80 bytecode-to-native-code compiler by
Deutsch and Schiffman~\cite{deutsch+:1984:efficient-implementation}, and the
optimizing \ac{JIT} compiler of Self, with type specialization and speculative
inlining~\cite{chambers+:1989:efficient-implementation}. These concepts were
later used in the HotSpot \ac{JIT} compiler~\cite{paleczny+:2001:java-hotspot}
for Java.

The prevalence of web browsers has made \ac{JIT} compilation an important topic
for JavaScript implementations, \eg the int V8 JavaScript
implementation~\cite{holtta:2013:crankshafting-from}. The map transitions for
hidden classes used in V8~\cite{google-inc:2012:chrome-documentation:} and
inspired by Self~\cite{chambers+:1989:efficient-implementation}, are in
principle similar to our notion of transformation rules. As well as objects in
V8 start with a default hidden class and follow map transitions to their most
optimal hidden class, the transformation rules in our approach change the shape
of a value object from its default shape to its most optimized one during the
value object's creation.

An important difference between the hidden classes of V8 to the shapes of our
approach is that V8 needs to deal with the objects being mutated after their
construction. Indeed, while the hidden classes of V8 (and similarly of
Higgs~\cite{chevalier-boisvert+:2015:extending-basic}) can encode the type of
the fields of the objects, they do that only for primitive values like int,
float etc. They cannot recursively express that a field is itself an object
with a specific hidden class, which is what we do with shapes in the current
paper. The reason this is impossible (or at least significantly harder) in the
JavaScript setting is the fact that the inner object can be mutated later,
which might cause its hidden class to change.

Tracing \ac{JIT} compilers as introduced by
Mitchell~\cite{mitchell:1970:design-construction} have seen implementations for
Java~\cite{gal+:2006:hotpathvm:-effective},
JavaScript~\cite{gal+:2009:trace-based-just-in-time}, or
Lua\footnote{\url{http://luajit.org}}, to name a few. In the context of a
JavaScript implementation, the SPUR
project~\cite{bebenita+:2010:spur:-trace-based} provided a tracing \ac{JIT}
compiler for Microsoft's \acf{CIL}.

Tracing an \emph{interpreter} that runs a program instead of tracing the
program itself it the core idea of meta-tracing \ac{JIT} compilers, pioneered
in the DynamoRIO project~\cite{sullivan+:2003:dynamic-native}.
PyPy~\cite{rigo+:2006:pypys-approach,bolz+:2009:tracing-meta-level:} is a
meta-circular Python implementation that uses a meta-tracing \ac{JIT} compiler.
Provided through the \RPython tool chain, other language implementations can
benefit from a meta-tracing, \eg Smalltalk~\cite{bolz+:2008:back-future},
Haskell~\cite{thomassen:2013:trace-based-just-in-time},
PHP\footnote{\url{http://hippyvm.com/}}, or
R\footnote{\url{https://bitbucket.org/roy_andrew/rapydo}}.%
The meta-tracing \ac{JIT} used in this work is provided by \RPython, as well.


\section{Conclusion and Future Work}
\label{sec:future-directions}

Our approach to just-in-time optimization of value classes provides very good
initial results both for execution time and memory consumption for a small
prototype implementation on selected micro-benchmarks. They are promising and
motivate us to investigate the matter further.

However, the current results are not yet fit for generalization. While our
prototypes give promising results on micro-benchmarks, they allow only limited
reasoning about more general programs. The applicability of our approach to
more general languages and especially more realistic programs remains to be
assessed in future work. Hence, immediate next steps include broadening the
benchmarks for the Pycket-based implementation so that we can assess the
viability of our approach in more representative context.

Racket supports more datatypes that may be subject to our approach, for example
(immutable) \emph{cons} cells. We plan to integrate these with our approach.

Our aim is then to broaden the scope of our approach beyond value classes. We
want to support objects that have identity as well as mutable objects. While
the usage of a cell indirection in the Pycket implementation has proven
worthwhile to allow mutability, we do not yet know whether this approach of
quasi-immutability is portable to other languages. Even more, maintaining
identity, and hence object-oriented concepts, needs more in-depth
investigation.

\section*{Acknowledgments}
\ifblind
(removed for blind review)\par
\vskip 2\baselineskip
\else
We gratefully acknowledge the financial support of HPI's Research School and
the Hasso Plattner Design Thinking Research Program (HPDTRP). Carl Friedrich
Bolz is supported by the EPSRC \emph{Cooler} grant EP/K01790X/1. We thank Alan
Borning for comments on a draft version of this paper. We thank the anonymous
reviewers for their detailed feedback.
\fi

\printbibliography
\appendix
\null

\begin{table}
  \caption[Execution times]{Benchmark execution times. We give means of the
    execution time along with
    the confidence interval showing the \SI{95}{\percent} confidence level.
  }
  \label{tab:time}
  \centering
  %latex.default(t, file = name, rowlabel = "Benchmark", rowlabel.just = "@{}l",     booktabs = TRUE, table.env = FALSE, center = "none", size = "scriptsize",     colheads = rep(c("mean", "error"), len), col.just = .just,     cgroup = levels(as.factor(ref$vm)), cdec = rep(0, len * 2))%
{\scriptsize
\begin{tabular}{@{}l@{}r@{}>{\smaller\ensuremath{\pm}}r@{\,\si{\milli\second}}cr@{}>{\smaller\ensuremath{\pm}}r@{\,\si{\milli\second}}cr@{}>{\smaller\ensuremath{\pm}}r@{\,\si{\milli\second}}cr@{}>{\smaller\ensuremath{\pm}}r@{\,\si{\milli\second}}cr@{}>{\smaller\ensuremath{\pm}}r@{\,\si{\milli\second}}}
\toprule
\multicolumn{1}{c}{\bfseries Benchmark}&\multicolumn{2}{c}{\bfseries Prototype}&\multicolumn{1}{c}{\bfseries }&\multicolumn{2}{c}{\bfseries Pycket (optimized)}&\multicolumn{1}{c}{\bfseries }&\multicolumn{2}{c}{\bfseries Pycket (original)}&\multicolumn{1}{c}{\bfseries }&\multicolumn{2}{c}{\bfseries Racket}&\multicolumn{1}{c}{\bfseries }&\multicolumn{2}{c}{\bfseries PyPy}\tabularnewline
\cline{2-3} \cline{5-6} \cline{8-9} \cline{11-12} \cline{14-15}
\multicolumn{1}{@{}l}{}&\multicolumn{1}{c}{mean}&\multicolumn{1}{c}{error}&\multicolumn{1}{c}{}&\multicolumn{1}{c}{mean}&\multicolumn{1}{c}{error}&\multicolumn{1}{c}{}&\multicolumn{1}{c}{mean}&\multicolumn{1}{c}{error}&\multicolumn{1}{c}{}&\multicolumn{1}{c}{mean}&\multicolumn{1}{c}{error}&\multicolumn{1}{c}{}&\multicolumn{1}{c}{mean}&\multicolumn{1}{c}{error}\tabularnewline
\midrule
append&$5088$&$27$&&$5545$&$32$&&$9432$&$52$&&$13218$&$ 42$&&$10655$&$32$\tabularnewline
filter&$1285$&$ 4$&&$4743$&$42$&&$5590$&$87$&&$14240$&$172$&&$ 6691$&$43$\tabularnewline
map&$6344$&$87$&&$5609$&$22$&&$8332$&$63$&&$15010$&$161$&&$ 9632$&$83$\tabularnewline
reverse&$ 350$&$ 6$&&$1172$&$ 5$&&$3347$&$45$&&$ 6421$&$159$&&$ 4864$&$27$\tabularnewline
tree&$2814$&$17$&&$2420$&$17$&&$3926$&$21$&&$ 3893$&$ 27$&&$ 7949$&$41$\tabularnewline
\bottomrule
\end{tabular}}

\end{table}

\begin{table}
  \caption[Memory consumption]{Benchmark memory consumption. We give means of the
    memory consumption along with
    the confidence interval showing the \SI{95}{\percent} confidence level.
  }
  \label{tab:mem}
  \centering
  %latex.default(t, file = name, rowlabel = "Benchmark", rowlabel.just = "@{}l",     booktabs = TRUE, table.env = FALSE, center = "none", size = "scriptsize",     colheads = rep(c("mean", "error"), len), col.just = .just,     cgroup = levels(as.factor(ref$vm)), cdec = rep(0, len * 2))%
{\scriptsize
\begin{tabular}{@{}l@{}r@{}>{\smaller\ensuremath{\pm}}r@{\,\si{\kilo\byte}}cr@{}>{\smaller\ensuremath{\pm}}r@{\,\si{\kilo\byte}}cr@{}>{\smaller\ensuremath{\pm}}r@{\,\si{\kilo\byte}}cr@{}>{\smaller\ensuremath{\pm}}r@{\,\si{\kilo\byte}}cr@{}>{\smaller\ensuremath{\pm}}r@{\,\si{\kilo\byte}}}
\toprule
\multicolumn{1}{c}{\bfseries Benchmark}&\multicolumn{2}{c}{\bfseries Prototype}&\multicolumn{1}{c}{\bfseries }&\multicolumn{2}{c}{\bfseries Pycket (optimized)}&\multicolumn{1}{c}{\bfseries }&\multicolumn{2}{c}{\bfseries Pycket (original)}&\multicolumn{1}{c}{\bfseries }&\multicolumn{2}{c}{\bfseries Racket}&\multicolumn{1}{c}{\bfseries }&\multicolumn{2}{c}{\bfseries PyPy}\tabularnewline
\cline{2-3} \cline{5-6} \cline{8-9} \cline{11-12} \cline{14-15}
\multicolumn{1}{@{}l}{}&\multicolumn{1}{c}{mean}&\multicolumn{1}{c}{error}&\multicolumn{1}{c}{}&\multicolumn{1}{c}{mean}&\multicolumn{1}{c}{error}&\multicolumn{1}{c}{}&\multicolumn{1}{c}{mean}&\multicolumn{1}{c}{error}&\multicolumn{1}{c}{}&\multicolumn{1}{c}{mean}&\multicolumn{1}{c}{error}&\multicolumn{1}{c}{}&\multicolumn{1}{c}{mean}&\multicolumn{1}{c}{error}\tabularnewline
\midrule
append&$1220256$&$0$&&$1826172$&$ 3$&&$3342362$&$2$&&$3393476$&$ 329$&&$4625309$&$2$\tabularnewline
filter&$ 522631$&$1$&&$1360879$&$ 3$&&$2195701$&$5$&&$2713104$&$5883$&&$2996973$&$9$\tabularnewline
map&$1141762$&$1$&&$1600280$&$ 1$&&$2709864$&$4$&&$2518191$&$9704$&&$3512922$&$2$\tabularnewline
reverse&$ 192552$&$1$&&$ 651634$&$ 8$&&$1604697$&$2$&&$1647545$&$1935$&&$3512605$&$5$\tabularnewline
tree&$  71233$&$1$&&$ 209180$&$55$&&$ 502130$&$2$&&$ 300500$&$  16$&&$ 956974$&$4$\tabularnewline
\bottomrule
\end{tabular}}

\end{table}

\end{document}